\documentclass[fleqn,12pt,twoside]{article}
\usepackage{espcrc1}
\usepackage{graphicx}

\newcommand{\AmS}{{\protect\the\textfont2
  A\kern-.1667em\lower.5ex\hbox{M}\kern-.125emS}}

\title{Coulomb effects in four nucleon continuum states}
\author{R. Lazauskas\address[LPSC]{Laboratoire de Physique Subatomique et de Cosmologie,
        53. avenue des Martyrs, 38026 Grenoble Cedex, France} \thanks{%
e-mail: lazauskas@lpsc.in2p3.fr}, J. Carbonell\addressmark{\tt}\thanks{%
e-mail: carbonell@lpsc.in2p3.fr}.}

\begin{document}

\maketitle

\begin{abstract}
The Faddeev-Yakubovski equations are solved in configuration space for low
energy four-nucleon continuum states. Coulomb interaction was included into
the formalism permitting an exact description of the scattering states in p+$%
^{3}$He and p+$^{3}$H systems.
\end{abstract}


\bigskip\bigskip Three- and four-nucleon systems are the testing ground for
studying the nuclear interaction. If the modern NN potentials have reached a
very high degree of accuracy in describing the two-nucleon data, they are
unable to account for the binding energies of the lightest nuclei. The use
of three-nucleon forces (3NF) is mandatory. By adjusting their parameters,
one can obtain a satisfactory description of the nuclear bound states up to
A=10 \cite{Pieper}.

Low energy three-nucleon scattering observables are quite insensitive to 3NF
effects. Furthermore, three-nucleon dynamics is relatively rigid once
deuteron and triton binding energies are fixed. The four nucleon continuum
states, containing sensible structures as thresholds and resonances, can
show a stronger dependence on the interaction models \cite{Fonseca}.

\smallskip We present in this contribution some results concerning low
energy four-nucleon continuum states. Faddeev-Yakubovski equations have been
modified to include Coulomb interactions and then solved in configuration
space. The n-$^{3}$H, p-$^{3}$He and p-$^{3}$H systems have been
investigated using MT I-III and several realistic NN potentials in
conjunction with Urbana IX (UIX) 3NF. Our scattering length results are
summarized in Table \ref{table:1}.

\begin{table}[tbh]
\caption{4N scattering lengths calculated using different interaction models.
}
\begin{tabular}{c||l|l||l|l||l|l|}
\cline{2-7}
& \multicolumn{2}{|c||}{MT I-III} & \multicolumn{2}{|c||}{Av. 14} &
\multicolumn{2}{|c|}{Av. 18+UIX} \\ \cline{2-7}
\multicolumn{1}{l||}{} & J$^{\pi }=0^{+}$ & J$^{\pi }=1^{+}$ & J$^{\pi
}=0^{+}$ & J$^{\pi }=1^{+}$ & J$^{\pi }=0^{+}$ & J$^{\pi }=1^{+}$ \\
\hline\hline
\multicolumn{1}{|l||}{n-$^{3}$H} & 4.10 & 3.63 & 4.28 & 3.81 & 4.04 & 3.60
\\ \hline
\multicolumn{1}{|l||}{p-$^{3}$He} & 11.5 & 9.20 & - & - & - & - \\ \hline
\multicolumn{1}{|l||}{p-$^{3}$H} & -63.1 & 5.50 & -13.9 & 5.77 & -16.5 & 5.39
\\ \hline
\end{tabular}
\label{table:1}
\end{table}

\begin{figure}[ht]
\vspace{-1.1cm}
\par
\begin{center}
\includegraphics[width=14.5cm]{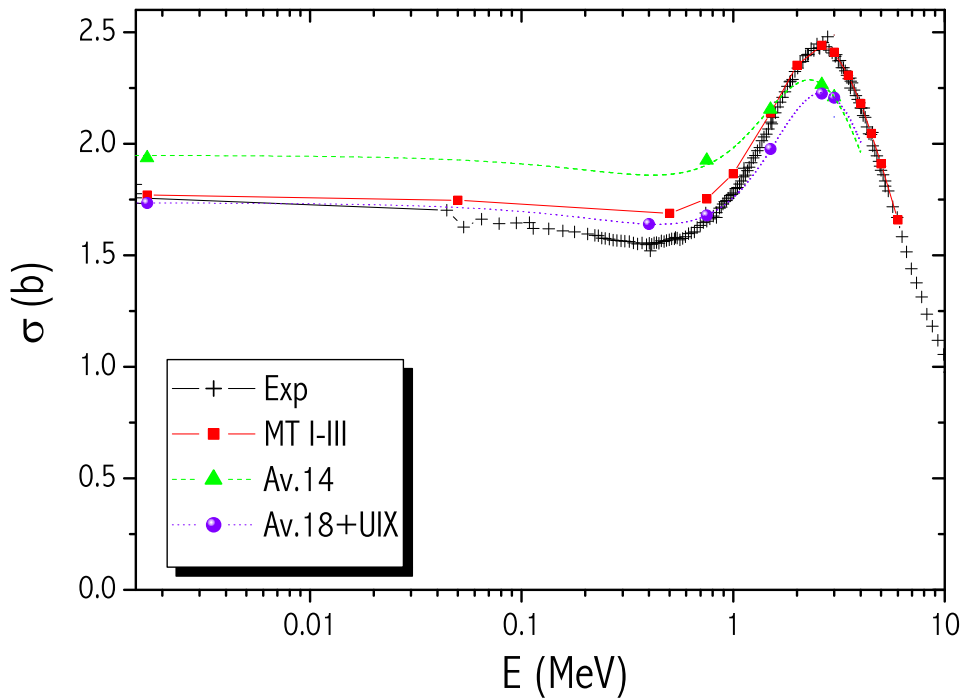}
\end{center}
\par
\vspace{-2.1cm}
\caption{Calculated n+$^{3}$H total cross sections compared with
experimental data \protect\cite{Phlips}.}
\label{Fig_nt_total_cs}
\end{figure}

The first effort was devoted to describe the Coulomb-free n-$^{3}$H system.
Semi-realistic MT I-III potential was shown to be very successful in
describing the total as well as the differential cross sections \cite
{Carbonell}. In our recent calculations with realistic Av.14 and Av.18
potentials, we were able to considerably enlarge the partial wave basis
(PWB) compared to \cite{Carbonell1}. Inspite of having some effect on the
negative parity phase shifts, the total cross section near the resonance
peak ($E_{cm}=3$ $MeV$) has not yet been improved (see Fig. \ref
{Fig_nt_total_cs}). We should notice, however, that the 2$^-$ phase shifts,
the most relevant contribution due to its statistical factor, are not yet fully
converged.
By including UIX-3NF we have managed to reproduce the
experimental zero-energy cross sections, which are overestimated by
realistic NN interaction without 3NF, but their effect near
the peak remains very small.

By including Coulomb interaction we were able to handle the n-$^{3}$H
isospin partner: p-$^{3}$He. Yet calculations were done with MT I-III model
only. The p-$^{3}$He scattering lengths predicted by this model (see Table
\ref{table:1}) agree with experimental values and are very close to the ones
obtained by the Pisa group \cite{Pisa}, using Av.18+UIX forces. Like in the
n-$^{3}$H case, MT I-III was found to be successful in describing
differential cross sections up to d+p+p threshold \cite{Thesis}.

The p-$^{3}$H scattering at energies below n-$^{3}$He threshold constitutes
a challenging problem due to the existence of a J$^{\pi}$=0$^{+}$, $^4$He
virtual state in between. The richness of this system makes scattering
observables very sensitive to the interaction and therefore provides an
excellent test of NN potentials. The splitting of p-$^{3}$H and n-$^{3}$He
thresholds is essentially due to Coulomb interactions. By properly taking
them into account in our calculations, we have placed the $^4$He virtual
state in between the two thresholds. This is reflected by a negative 0$^{+}$
p-$^{3}$H scattering length. Note that in all the preceding works, where
Coulomb interaction was neglected, the first $^{4}$He excitation  appeared
as a bound state.

Unlike in the other 4N systems, MT I-III predictions for p-$^{3}$H
scattering lengths as well as the excitation function -- $\left.{\frac{%
d\sigma}{d\Omega}}(E)\right|_{\theta=120^{\circ}}$ -- are in disagreement
with experimental data. $^4$He virtual state is located too close to the p-$%
^{3}$H threshold and has very small width.

\begin{figure}[ht]
\begin{center}
\vspace{-1.1cm}\includegraphics[width=13.5cm]{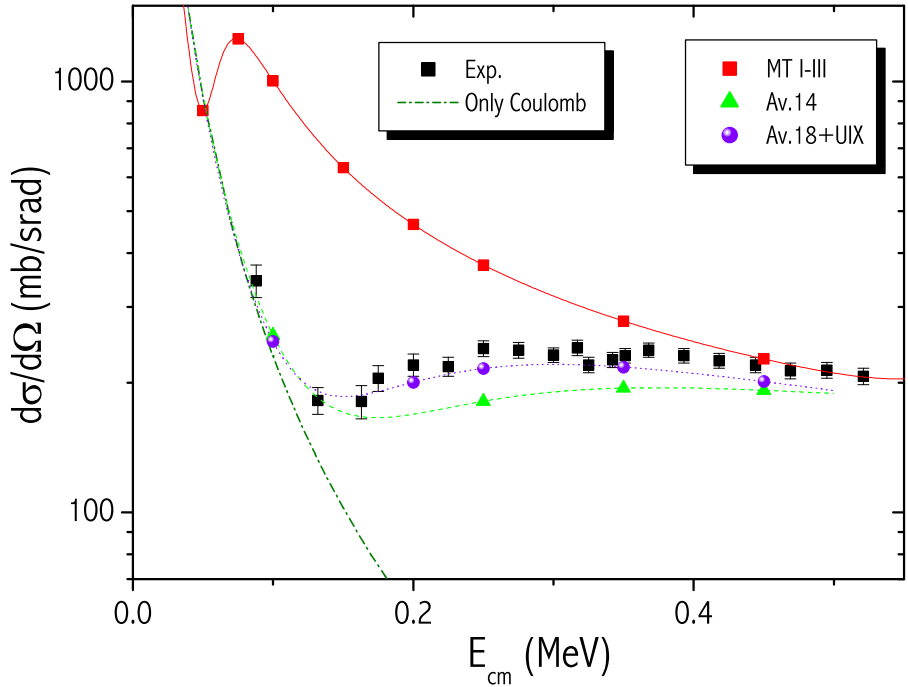}
\vspace{-1.5cm}
\end{center}
\caption{Energy dependence of p+$^{3}$He elastic differential cross sections
at 120$^{\circ}$: experimental results are compared with our calculation.}
\label{Fig_pt_cross_av}
\end{figure}

Our calculations with realistic potentials are still limited in PWB.
Nevertheless, they provide the very promissing results displayed in Fig. \ref
{Fig_pt_cross_av}. Pure 2NF models predict too large singlet scattering
length, thus placing the virtual state too far from the threshold. On the
other hand by implementing UIX-3NF in conjunction with Av.18 NN model one
obtains singlet scattering length as well as the excitation function $\left.{%
\frac{d\sigma}{d\Omega}}(E)\right|_{\theta=120^{\circ}}$ in agreement with
experimental data \cite{Pt_exp}. A detailed analysis of these calculations
is in progress \cite{Thesis,LC_JSPQ_03} as well as their extension above the
n-3He threshold.

\end{document}